\documentclass[twocolumn,english,aps,pra,showpacs,amsfonts,amsmath,amssymb,superscriptaddress,footinbib,floatfix,longbibliography]{revtex4-1}
\pdfoutput=1
\usepackage[T1]{fontenc}
\usepackage[latin9]{inputenc}
\usepackage{mathtools}
\usepackage{amsmath}
\usepackage{amssymb}
\usepackage{graphicx}
\usepackage{lipsum}
\usepackage{siunitx}
\usepackage{braket}
\makeatletter
\usepackage{ulem}
\usepackage{color}
\usepackage{filemod}

\newcommand{\hidetxt}[1]{}

\newcommand*\diff{\mathop{}\!\mathrm{d}}

\usepackage{babel}

\makeatother

\usepackage{babel}

\usepackage{hyperref}

\begin{document}

\title{Blockade-induced resonant enhancement of the optical nonlinearity in a Rydberg medium}

\author{Annika Tebben}

\affiliation{Physikalisches Institut, Universit\"at Heidelberg, Im Neuenheimer Feld 226, 69120 Heidelberg, Germany}
\author{Cl\'{e}ment Hainaut}
\affiliation{Physikalisches Institut, Universit\"at Heidelberg, Im Neuenheimer Feld 226, 69120 Heidelberg, Germany}

\author{Valentin Walther}
\affiliation{Department of Physics and Astronomy, Aarhus University, Ny Munkegade 120, DK 8000 Aarhus C, Denmark}

\author{Yong-Chang Zhang}
\affiliation{Department of Physics and Astronomy, Aarhus University, Ny Munkegade 120, DK 8000 Aarhus C, Denmark}

\author{Gerhard Z\"urn}
\affiliation{Physikalisches Institut, Universit\"at Heidelberg, Im Neuenheimer Feld 226, 69120 Heidelberg, Germany}
 
\author{Thomas Pohl}
\affiliation{Department of Physics and Astronomy, Aarhus University, Ny Munkegade 120, DK 8000 Aarhus C, Denmark}

\author{Matthias Weidem\"uller}
\affiliation{Physikalisches Institut, Universit\"at Heidelberg, Im Neuenheimer Feld 226, 69120 Heidelberg, Germany}
\affiliation{National Laboratory for Physical Sciences at Microscale and Department of Modern Physics, and CAS Center for Excellence and Synergetic Innovation Center in Quantum Information and Quantum Physics, Shanghai Branch, University of Science and Technology of China, Shanghai 201315, China}

\begin{abstract}
We predict a resonant enhancement of the nonlinear optical response of an interacting Rydberg gas under conditions of electromagnetically induced transparency. The enhancement originates from a two-photon process which resonantly couples electronic states of a pair of atoms dressed by a strong control field. We calculate the optical response for the three-level system by explicitly including the dynamics of the intermediate state.  We find an analytical expression for the third order susceptibility for a weak classical probe field. The nonlinear absorption displays the strongest resonant behavior on two-photon resonance where the detuning of the probe field equals the Rabi frequency of the control field. The nonlinear dispersion of the medium exhibits various spatial shapes depending on the interaction strength. Based on the developed model, we propose a realistic experimental scenario to observe the resonance by performing transmission measurements.

\end{abstract}

\date{\today}

\maketitle
 
\section{Introduction}
\label{sec:intro}
A Rydberg gas under conditions of electromagnetically induced transparency (EIT) exhibits a nonlinear optical response, which exceeds that of conventional media by orders of magnitude \cite{FirstenbergAdamsHofferberth:review:JPhysB2016,Pohl:review:16} . Aiming at the full control of effective photon interactions, numerous experimental achievements, such as the realization of single-photon transistors \cite{Gorniaczyk:PhotonTransistor:PRL2014,Tiarks:PhotonTransistor:PRL2014,Tiarks:PhotonTransistor:PRL2014} and the creation of bound states of photons \cite{Firstenberg:AttractivePhotons:Nature2013,Liang:ThreePhoton:Science2018}, as well as advanced theoretical investigations both in the quantum \cite{Fleischhauer:DarkStatePolariton:PRL00, Fleischhauer:QuantumMemory:PRA02,Gorshkov:PhotonInteractionsBlockade:PRL11,Bienias:ScatteringResonances:PRA2014,Gullans:EffectiveFieldTheory:PRL2016} and semi-classical regimes \cite{Ates:EITuniversality:PRA2011,Gaerttner:SemianalyticalModel:PRA14,Sevincli:Adams:CPTEIT:JoPB11,Sevincli:adibatic:PRL11} have been reported.

In the quantum regime, the notion of dark-state polaritons has proven to be successful for the theoretical description of photon propagation through an interacting Rydberg medium \cite{Fleischhauer:DarkStatePolariton:PRL00, Fleischhauer:QuantumMemory:PRA02}. In the case of two interacting photons a wavefunction approach was developed \cite{Gorshkov:PhotonInteractionsBlockade:PRL11} and allowed to accurately describe the experimental findings of dissipative \cite{Peyronel:DissipativeQuantum:Nature2012}, spin-exchange like \cite{Thompson:SymmetryProtectedCollisions:Nature2017} as well as attractive photonic interactions \cite{Firstenberg:AttractivePhotons:Nature2013,Liang:ThreePhoton:Science2018}. More complex models to describe the photon propagation investigated the scattering properties of two polaritons \cite{Bienias:ScatteringResonances:PRA2014} and made the transition to the few- and many-body regime in one dimension utilizing an effective field theory \cite{Gullans:EffectiveFieldTheory:PRL2016}. 

In the semi-classical regime, a Monte Carlo rate equation model was used to obtain an expression for the nonlinear response of the atomic gas by including Rydberg interactions as level shifts \cite{Ates:EITuniversality:PRA2011} or by using a superatom approach \cite{Gaerttner:SemianalyticalModel:PRA14}. This picture was condensed to a universal scaling of the nonlinear absorption with the fraction of Rydberg blockaded atoms. This scaling proved to be consistent with calculations in the quantum regime, that are typically much more complicated. Moreover, it showed excellent agreement with experimental results \cite{Sevincli:Adams:CPTEIT:JoPB11}, underlining the strength of this basic model. 
Due to the long-range interactions between Rydberg atoms, the nonlinearity in Rydberg-EIT systems is intrinsically nonlocal.
Based on a cluster expansion, an analytic expression for this nonlocal optical response of a Rydberg gas has been derived \cite{Sevincli:adibatic:PRL11}. These results proved the existence of modulational instabilities, which are a precursor of photon crystallization. 
All these semi-classical approaches neglect the dynamics of the intermediate state. However, including these dynamics revealed interesting characteristics of the photonic and atomic pair potentials \cite{Bienias:ScatteringResonances:PRA2014,Gaul:ResDressing:PRL16,Helmrich:ResDressing:PRL16}.

\begin{figure}[t!]
	\includegraphics[width=1\linewidth]{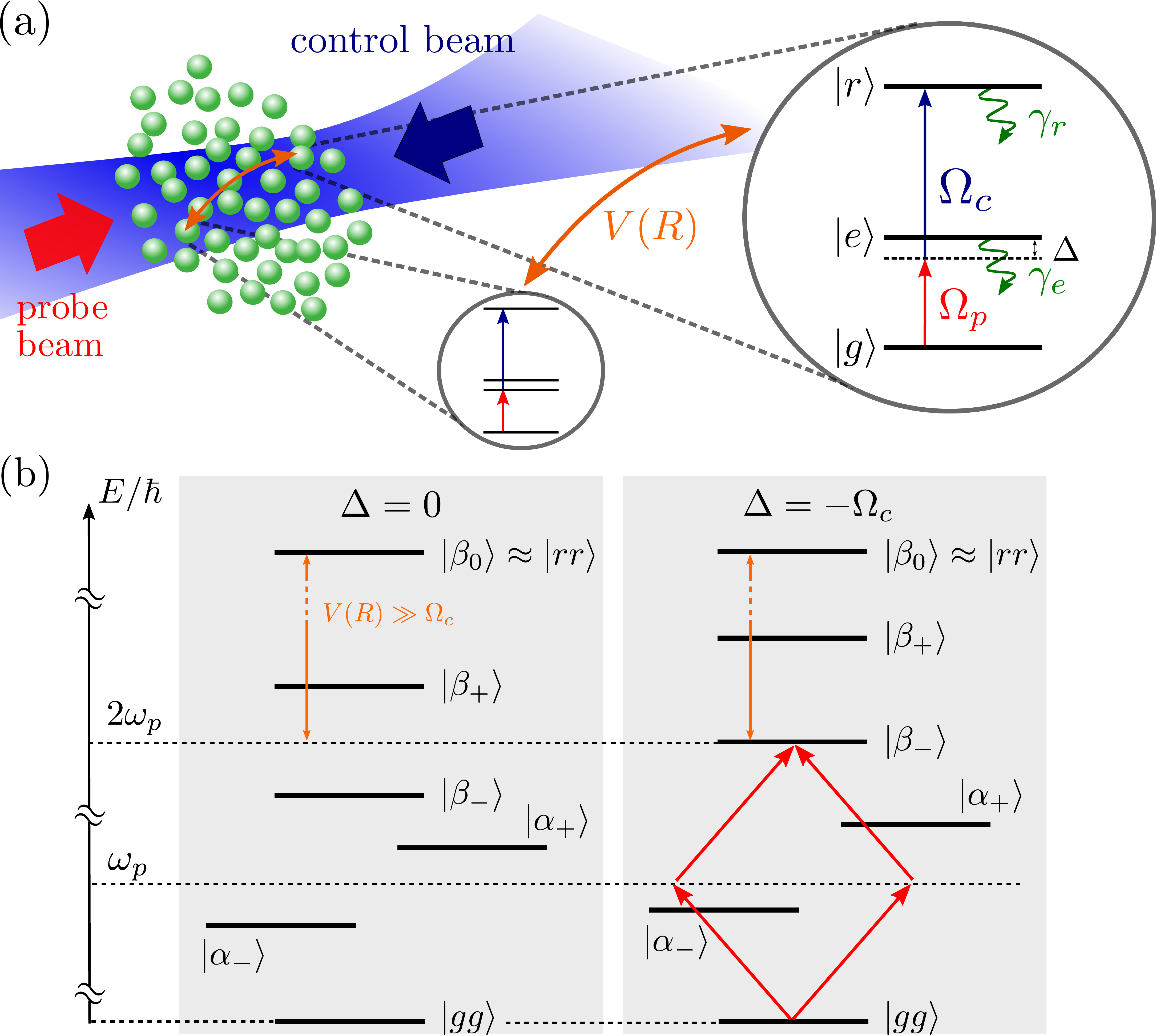}\caption{(a) Atomic level structure of Rydberg atoms, which interact via an interaction $V(R)$, that depends on the inter-atomic distance $R$. (b) Relevant level scheme in the dressed pair-state basis, as explained in the main text. For $\Delta=-\Omega_c$ the eigenstate $\ket{\beta_-}$ moves into two-photon resonance with the ground state $\ket{gg}$.}
	\label{fig:level-scheme_resonance} 
\end{figure}

Here, we develop a semi-classical model for the nonlocal, nonlinear response
of an interacting Rydberg gas, explicitly including the dynamics of the intermediate state. We reveal the existence of a two-body, two-photon resonance in the optical response when the control field Rabi frequency is tuned to the probe field detuning.

In order to provide a simple picture, we start by describing the system based on a pair-state model and explain how atomic interactions lead to a two-photon resonance. We then derive an analytical expression for the nonlinear response of the interacting Rydberg gas for arbitrary interaction strengths starting from the Maxwell-Bloch equations. We show that in the presence of the resonance the nonlinear response can be significantly enhanced. We discuss the spatially dependent absorption features of the nonlinear response and present the scaling of the enhancement with relevant field and atom parameters. Finally, we propose a feasible transmission measurement revealing the resonance.

\section{Laser-dressed interacting pair-states}
\label{sec:two_body}
Consider a ladder-type realization of the EIT scheme, where a gas of Rydberg atoms with density $\rho$ is exposed to counter-propagating probe and control fields as shown in Fig. \ref{fig:level-scheme_resonance}(a). The coherent probe field $\mathcal{E}(\mathbf{r},t)$ with frequency $\omega_p$ and Rabi frequency $\Omega_p$ couples the atomic ground state $\ket{g}$ to a short lived intermediate state $\ket{e}$ with decay rate $\gamma_e$, while a control field with Rabi frequency $\Omega_c$ drives the transition to a metastable Rydberg state $\ket{r}$ with a small decay rate $\gamma_r$. The two-photon detuning $\delta$ for the ground to Rydberg state transition is kept at zero, but the fields are detuned from the intermediate state by the single-photon detuning $\Delta$, as shown in Fig. \ref{fig:level-scheme_resonance}(a).

The system is governed by pairwise van der Waals interactions $V(R)$, giving rise to the so-called Rydberg blockade effect. Here, two atoms at a distance smaller than the blockade radius $R_b$ cannot simultaneously be excited to the Rydberg states \cite{Lukin:RydbergBlockade:RPL01}. In the case of Rydberg EIT, $R_b$ is defined as the distance where the van der Waals potential exceeds the EIT linewidth $\delta_\text{EIT}=\Omega_c^2/\lvert\gamma_e-i\Delta\rvert$ \cite{Fleischhauer:EIT:PMP05}. Thus, $R_b=(c_6/\delta_\text{EIT})^{1/6}$ is the characteristic length scale of the system.

Considering pair-wise interactions, it is natural to examine the coupled atom-light system in the pair-state basis. 
The corresponding Hamiltonian \cite{Gaul:ResDressing:PRL16}
\begin{equation}
\hat{\mathcal{H}}=
\begin{pmatrix}
0 & \sqrt{2}\Omega_p & 0 & 0 & 0 & 0\\
\sqrt{2}\Omega_p & -\Delta & \Omega_c & \sqrt{2}\Omega_p & 0& 0\\
0 & \Omega_c & 0 & 0 & \Omega_p& 0\\
0 & \sqrt{2}\Omega_p & 0 & -2\Delta  & \sqrt{2}\Omega_c& 0\\
0 & 0 & \Omega_p  & \sqrt{2}\Omega_c & -\Delta& \sqrt{2}\Omega_c\\
0 & 0 & 0  & 0 &\sqrt{2}\Omega_c & V(R) \\
\end{pmatrix} 
\label{equ:H_pair-states}
\end{equation}
describes the coupling between the ground state $\ket{gg}$ and the states $\{\ket{ge}_+,\ket{gr}_+\}$ and $\{\ket{ee},\ket{er}_+,\ket{rr}\}$ in the singly- and doubly excited subspaces, respectively. Here, we make use of the symmetric pair-state basis, where   $\ket{ij}_+=(\ket{ij}+\ket{ji})/\sqrt{2}$ with $i,j\in \{ g,e,r \}$. 

In the limit of vanishing interactions ($V(R)\rightarrow 0$) at large interatomic distances, the system reduces to a gas of individual atoms under EIT conditions, featuring a linear response to the applied fields \cite{Fleischhauer:EIT:PMP05}.

In the following, we discuss how the presence of interactions changes the energy spectrum of the eigenstates of $\hat{\mathcal{H}}$. For $\Omega_c\gg\Omega_p$ the singly- and doubly excited subspaces can be dressed by the control field individually \cite{Gaul:ResDressing:PRL16}, leading to eigenstates $\{\ket{\alpha}_+,\ket{\alpha}_-\}$ and $\{\ket{\beta}_-,\ket{\beta}_+,\ket{\beta}_0\}$, respectively, as shown in Fig. \ref{fig:level-scheme_resonance}(b).

\begin{figure}[t!]
	\includegraphics[width=1\linewidth]{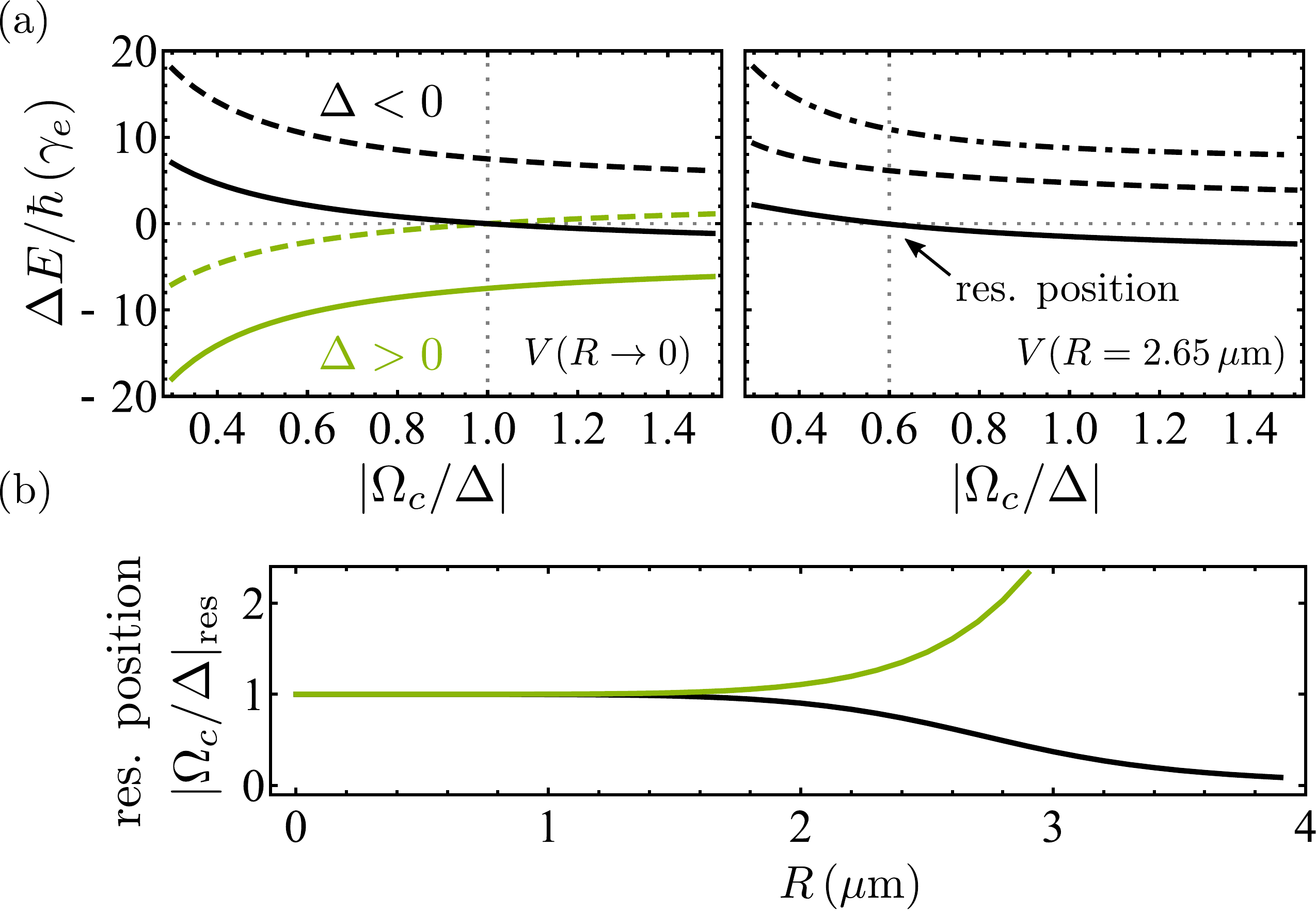}
	\caption{ (a) Energy $\Delta E$ of the dressed levels $\ket{\beta_-}$ (solid line), $\ket{\beta_+}$ (dashed line) and $\ket{\beta_0}$ (dashed-dotted) against the ratio $\lvert\Omega_c/\Delta\rvert$ for positive (green) and negative values (black) of $\Delta$, respectively. $\Delta E=0$ corresponds to the two-photon resonance with the ground state, which is met for infinite ($R\rightarrow0$, left) and finite (right) interactions for different $\lvert\Omega_c/\Delta\rvert$. (b) Resonance position $\lvert\Omega_c/\Delta\rvert_\text{res}$ against the inter-atomic distance $R$ for positive (green) and negative (black) single-photon detunings. }
	\label{fig:pair-states_energies} 
\end{figure}

In the limit of strong interactions ($V(R)\gg\Omega_c$), the eigenstate $\ket{\beta_0}$ mainly contains the doubly excited Rydberg state and is decoupled from the remaining level system, as schematically shown in Fig. \ref{fig:level-scheme_resonance}(b). Here, the ground state $\ket{gg}$ is coupled by two probe photons to the dressed states of the doubly-excited subspace. This coupling becomes maximal for $\Omega_c=\pm\Delta$, as shown in Fig. \ref{fig:pair-states_energies}(a), and establishes a two-body, two-photon resonance, that has already inspired the method of resonant Rydberg dressing \cite{Gaul:ResDressing:PRL16,Helmrich:ResDressing:PRL16}. 

In the case of finite interactions, the  influence of the doubly excited Rydberg state $\ket{rr}$ on the energy spectrum has to be considered explicitly. Here, the dressed state $\ket{\beta}_0$ alters the energy spectrum and shifts the states $\ket{\beta_\pm}$ to lower energies as shown exemplarily for $V(R=\SI{2.64}{\mu m})$ and $\Delta<0$ in the right graph of Fig. \ref{fig:pair-states_energies}(a). For $\Delta>0$ this happens in a similar manner, such that we only display one case here for clarity. As a result of these energy shifts, the ratio where the two-photon resonance condition is met shifts to smaller values, in this example to $\lvert\Omega_c/\Delta\rvert=0.6$.

Fig. \ref{fig:pair-states_energies}(b) highlights this effect and shows the resonance position $\lvert\Omega_c/\Delta\rvert_\text{res}$ against the inter-atomic separation $R$, meaning different interaction strengths. For every $R$ there exists exactly one ratio $\lvert\Omega_c/\Delta\rvert$ for negative (black) and positive (green) single-photon detunings, where the resonance condition is met.  

Considering the propagation of the probe field, this resonance changes the nonlinear optical response of the Rydberg gas, for which we will derive an analytical expression in the following.

\section{Nonlinear optical response}
\label{sec:cluster-expansion}
In this section, we derive a spatially dependent analytical expression for the nonlinear, nonlocal susceptibility of the Rydberg EIT gas, that allows to study the optical response for various interaction strengths, non-flat probe fields and non-constant atomic density distributions. For this purpose, we first introduce a set of bosonic Maxwell-Bloch equations that accurately describe the interacting many-body system under weak-driving conditions. Next, we proceed by solving these equations for a classical probe field exactly up to the third order in a cluster expansion. Finally, we discuss the spatially-dependent refraction and absorption features of the nonlinear, nonlocal susceptibility.

\subsection{Maxwell-Bloch equations}
The bosonic Maxwell-Bloch equations for the Rydberg-EIT system read \cite{Pohl:review:16}
\begin{align}
\partial_t \mathcal{E}(\mathbf{r})=
&\left(ic\frac{\nabla^2_{\perp}}{2k_p}-c\partial_z\right)\mathcal{E}(\mathbf{r})
-ig\sqrt{\rho}\hat{P}(\mathbf{r}),
\label{equ:Max-Bloch_int_1}\\
\partial_t \hat{P}(\mathbf{r})= 
&-ig\sqrt{\rho}\mathcal{E}(\mathbf{r})
-i\Omega_c(\mathbf{r})\hat{S}(\mathbf{r})
-\Gamma_e\hat{P}(\mathbf{r}),
\label{equ:Max-Bloch_int_2}\\
\partial_t \hat{S}(\mathbf{r})=
&-i\Omega_c(\mathbf{r})\hat{P}(\mathbf{r})
-\Gamma_r\hat{S}(\mathbf{r})\nonumber\\
&-i\int\diff\mathbf{r'}V(\mathbf{r}-\mathbf{r'})\hat{S}^{\dagger}(\mathbf{r'})\hat{S}(\mathbf{r'})\hat{S}(\mathbf{r}) \, ,
\label{equ:Max-Bloch_int_3}
\end{align}
where we dropped the time-dependence of the fields and operators for convenience.

Eq. (\ref{equ:Max-Bloch_int_1}) describes, in paraxial approximation, the propagation of a classical probe field $\mathcal{E}(\mathbf{r},t)$ in $z$-direction through a medium with source term $-ig\sqrt{\rho}\hat{P}(\mathbf{r})$. Here, $g\sqrt{\rho}$ is the collectively enhanced single-atom coupling strength $g$ of the probe transition, $k_p=\omega_p/c$ the wavenumber of the probe field, $c$ the speed of light, and $\hat{P}(\mathbf{r})$ a bosonic operator for the polarisation coherence as motivated below. The assumption of a classical probe field is meaningful, if photon-photon and photon-atom correlations can be neglected, implying that  the coherent nature of the field is preserved \cite{Pohl:review:16}. This is true, as long as the atomic interactions and the coupling $g$ to the probe field are small. In the case of Rydberg-EIT, this is given for an optical depth per blockade radius $\text{OD}_b\propto g\rho c_6^{1/6}\ll 1$ \cite{Pohl:review:16}.

If the probe field is weak compared to the control field, the atomic part of the Maxwell-Bloch equations is reasonably described in terms of continuous bosonic operators $\hat{P}(\mathbf{r},t)$ and $\hat{S}(\mathbf{r},t)$ for the polarisation and Rydberg spin-wave coherence, respectively \cite{Fleischhauer:DarkStatePolariton:PRL00,Fleischhauer:QuantumMemory:PRA02}. Moreover, within the weak-probe assumption, population decay can be neglected and only the coherence decay rates $\bar{\gamma}_{e,r}=\gamma_{e,r}/2$ remain. We defined $\Gamma_e = \bar{\gamma}_e-i\Delta$ and $\Gamma_r = \bar{\gamma}_r-i\delta$.

Eq. (\ref{equ:Max-Bloch_int_1}) to (\ref{equ:Max-Bloch_int_3}) have been solved  in the semi-classical regime for $\Delta$ or $\gamma_e\gg\Omega_c$ \cite{Sevincli:adibatic:PRL11,Pohl:review:16}, where the intermediate state dynamics can be eliminated. In these works, it has been shown, that the Rydberg EIT system exhibits a strong nonlinear and nonlocal response to the driving field. Motivated by this, we recast, in steady-sate, Eq. (\ref{equ:Max-Bloch_int_1}) into 
\begin{align}
i\partial_z \mathcal{E}(\mathbf{r})=
&-\frac{\nabla^2_{\perp}}{2k_p}\mathcal{E}(\mathbf{r})
+\chi^{(1)}(\mathbf{r})\mathcal{E}(\mathbf{r})\nonumber\\
&+\int \diff \mathbf{r'} 
\chi^{(3)}(\mathbf{r}-\mathbf{r'})\lvert\mathcal{E}(\mathbf{r'})\rvert^2\mathcal{E}(\mathbf{r})\, ,
\label{equ:propagation_E}
\end{align}
where the linear $\chi^{(1)}(\mathbf{r})$ and nonlinear susceptibility $\chi^{(3)}(\mathbf{r}-\mathbf{r'})$ are directly related to the polarisation coherence via
\begin{align}
\langle\hat{P}(\mathbf{r})\rangle =&\frac{c}{g\sqrt{\rho}}\left[ \chi^{(1)}(\mathbf{r})\mathcal{E}(\mathbf{r})\right.\nonumber\\
&\left.+\int \diff \mathbf{r'} 
\chi^{(3)}(\mathbf{r}-\mathbf{r'})\lvert\mathcal{E}(\mathbf{r'})\rvert^2\mathcal{E}(\mathbf{r})\right]\, .
\label{equ:polarisation_susceptibility_relation}
\end{align}
In Eq. (\ref{equ:propagation_E}), the two complex susceptibilities given in Eq. (\ref{equ:chi1}) and (\ref{equ:chi3}) act as an effective light potential responsible for refraction and absorption on the linear and nonlinear level, respectively.

\subsection{Perturbative solution}
For $\Omega_p\ll\Omega_c$, we proceed by solving the Maxwell-Bloch equations with a perturbative expansion in the probe field. For this purpose we separate the probe field as $\mathcal{E}(\mathbf{r})=\mathcal{E}_0f(\mathbf{r})$, where the position dependence is absorbed in $f(\mathbf{r})$ and $\mathcal{E}_0$ is a small parameter. We expand the expectation values of the polarisation coherence in terms of $\mathcal{E}_0$ as
\begin{align}
\langle \hat{P}(\mathbf{r})\rangle=P^{(0)}(\mathbf{r}) &+ \mathcal{E}_0P^{(1)}(\mathbf{r})+\mathcal{E}_0^2P^{(2)}(\mathbf{r})\nonumber\\
&+\mathcal{E}_0^3P^{(3)}(\mathbf{r})+\mathcal{O}(\mathcal{E}_0^4)
\end{align}
and similarly for the spin-wave coherence $\hat{S}(\mathbf{r})$. Inserting this into Eq. (\ref{equ:Max-Bloch_int_2}) and (\ref{equ:Max-Bloch_int_3}) allows to solve the problem order by order.

In zeroth-order the probe field vanishes, such that all atoms remain in the ground state. Therefore, $\mathcal{P}^{(0)}(\mathbf{r})=\mathcal{S}^{(0)}(\mathbf{r})=0$. Moreover, the second- and all higher even orders vanish due to the centro-symmetry of the atomic gas.

The first-order has the solution
\begin{align}
P^{(1)}(\mathbf{r})&=-ig\sqrt{\rho}\frac{\Gamma_r}{\Omega^2(\mathbf{r})+\Gamma_r\Gamma_e}f(\mathbf{r})
\label{equ:P_1},\\
S^{(1)}(\mathbf{r})&=-g\sqrt{\rho}\frac{\Omega(\mathbf{r})}{\Omega^2(\mathbf{r})+\Gamma_r\Gamma_e}f(\mathbf{r})\, .
\label{equ:S_1}
\end{align}
Inserting the result for $P^{(1)}(\mathbf{r})$  into Eq. (\ref{equ:polarisation_susceptibility_relation}) leads to the linear susceptibility
\begin{equation}
\chi^{(1)}(\mathbf{r})=-ig^2\frac{\Gamma_r}{c(\Omega_c^2+\Gamma_r\Gamma_e)}\rho(\mathbf{r})\, .
\label{equ:chi1}
\end{equation}
It recovers the well-known effect of EIT in the absence of atomic interactions and leads, for $\gamma_r=0$, to a full transmission of the probe field on two-photon resonance ($\delta=0$).

Solving the third-order equations
\begin{align}
\partial_t P^{(3)}(\mathbf{r}) = &-i\Omega(\mathbf{r})S^{(3)}(\mathbf{r})-\Gamma_e P^{(3)}(\mathbf{r})
\label{equ:P1}, \\
\partial_t S^{(3)}(\mathbf{r})= &-i\Omega(\mathbf{r})P^{(3)}(\mathbf{r})-\Gamma_rS^{(3)}(\mathbf{r})\nonumber\\
&-i\int \diff\mathbf{r'}V(\mathbf{r}-\mathbf{r'})
\langle \hat{S}^{\dagger}(\mathbf{r'})\hat{S}(\mathbf{r'})\hat{S}(\mathbf{r}) \rangle
\end{align}
is more involved due to the appearance of correlations between Rydberg spin-wave excitations $ \langle\hat{S}^\dagger(\mathbf{r'})\hat{S}(\mathbf{r'})\hat{S}(\mathbf{r})\rangle$. In the following, we explain the main steps of calculating this correlator.

The time dependence of the Rydberg spin wave correlator
\begin{align}
\partial_t &\langle\hat{S}^\dagger(\mathbf{r'})\hat{S}(\mathbf{r'})\hat{S}(\mathbf{r})\rangle=\nonumber\\
& -i\Omega_c(\mathbf{r})\,\langle\hat{S}^\dagger(\mathbf{r'})\hat{S}(\mathbf{r'})P(\mathbf{r})\rangle\nonumber\\
&+i\Omega_c(\mathbf{r'})\left[\langle\hat{P}^\dagger(\mathbf{r'})\hat{S}(\mathbf{r'})\hat{S}(\mathbf{r})\rangle-\langle\hat{S}^\dagger(\mathbf{r'})\hat{P}(\mathbf{r'})\hat{S}(\mathbf{r})\rangle \right]\nonumber\\
&-\left[3\bar{\gamma}_r+iV(\mathbf{r}-\mathbf{r'})\right]\langle\hat{S}^\dagger(\mathbf{r'})\hat{S}(\mathbf{r'})\hat{S}(\mathbf{r})\rangle\nonumber\\
&-i\int\diff\mathbf{r''}V(\mathbf{r'}-\mathbf{r''})\langle\hat{S}^\dagger(\mathbf{r'}) \hat{S}^\dagger(\mathbf{r''}) \hat{S}(\mathbf{r'})\hat{S}(\mathbf{r''})\hat{S}(\mathbf{r})\rangle
\label{equ:SSS_timeDerivative}
\end{align}
is given by Eq. (\ref{equ:Max-Bloch_int_2}) and (\ref{equ:Max-Bloch_int_3}).
As a two-body correlator it requires knowledge of other two-body correlators as for instance $\langle\hat{P}^\dagger(\mathbf{r'})\hat{S}(\mathbf{r'})\hat{S}(\mathbf{r})\rangle$, as well as the three-body correlator in the last line of Eq. (\ref{equ:SSS_timeDerivative}). Ultimately this leads to an infinite hierarchy of equations for the many-body system, that needs to be truncated appropriately.

Here, the weak-probe assumption in combination with the blockade effect provides a natural way of truncating the hierarchy as it limits the density of Rydberg excitations in the system \cite{Pohl:review:16}. Therefore, the probability of finding two Rydberg excitations within a blockaded volume is small, and becomes negligible for three or more excitations. In this case, we can discard three-body interactions and correlations of this and higher orders are fully suppressed \cite{Pohl:review:16}. This does not only allow to truncate the hierarchy of equations, but also implies that two-body atomic correlations are taken into account exactly.

Applying this approach, we neglect terms as for example the last line of Eq. (\ref{equ:SSS_timeDerivative}) and in a similar manner obtain the time derivatives of all involved one- and two-body correlators. This leads to 20 coupled, linear equations. In order to proceed with the calculation, we make the ansatz
\begin{align}
\hat{P}(\mathbf{r}) &= -ig\sqrt{\rho}\mathcal{E}_0f(\mathbf{r})\frac{\Gamma_r}{\Omega^2(\mathbf{r})+\Gamma_r\Gamma_e}\hat{P'}(\mathbf{r}),\\
\hat{S}(\mathbf{r})  &= -g\sqrt{\rho}\mathcal{E}_0f(\mathbf{r})\frac{\Omega(\mathbf{r})}{\Omega^2(\mathbf{r})+\Gamma_r\Gamma_e}\hat{S'}(\mathbf{r}) \, .
\end{align}
motivated by the first-order solutions of the Maxwell-Bloch equations given in Eq. (\ref{equ:P_1}) and (\ref{equ:S_1}). Performing the associated variable change and assuming a spatially constant Rabi frequency of the control field, makes the equations position-independent and allows to rephrase them as a $20\times 20$-matrix in the steady-state. Solving the system gives the exact solution for the Rydberg spin-wave correlator
\begin{align}
\langle&\hat{S}^{\dagger}(\mathbf{r'})\hat{S}(\mathbf{r'})\hat{S}(\mathbf{r})\rangle=\nonumber\\
&-\frac{\Omega_c^3}{\lvert a\rvert^2}
\frac{2(\Gamma_r+\Gamma_e) g^3\rho(\mathbf{r'})\sqrt{\rho(\mathbf{r})}}{2a(\Gamma_r+\Gamma_e)+i(a+\Gamma_e^2)V(\mathbf{r}-\mathbf{r'})}
\lvert \mathcal{E}(\mathbf{r'})\rvert^2 \mathcal{E}(\mathbf{r})
\end{align}
up to two-body interactions,
where we introduced the abbreviation $a=\Omega_c^2+\Gamma_r\Gamma_e$. Inserting this expression in the third-order equations of the expansion finally leads to the third-order susceptibility 
\begin{align}
\chi^{(3)}(\mathbf{r}-\mathbf{r'})= 
&\frac{\Omega_c^4g^4\rho(\mathbf{r'})\rho(\mathbf{r})}{c\lvert a\rvert^2 a}\nonumber\\
&\times\frac{2(\Gamma_r+\Gamma_e)V(\mathbf{r}-\mathbf{r'})}{2a(\Gamma_r+\Gamma_e)+i(a+\Gamma_e^2)V(\mathbf{r}-\mathbf{r'})}.
\label{equ:chi3}
\end{align}
Having obtained a result for the first- and third-order susceptibility we arrive at a closed Eq. (\ref{equ:propagation_E}) for the propagation of the probe field through the highly nonlinear and nonlocal Rydberg EIT medium.

\subsection{Spatial shape of the nonlinearity}
\begin{figure}[t!]
	\includegraphics[width=1\linewidth]{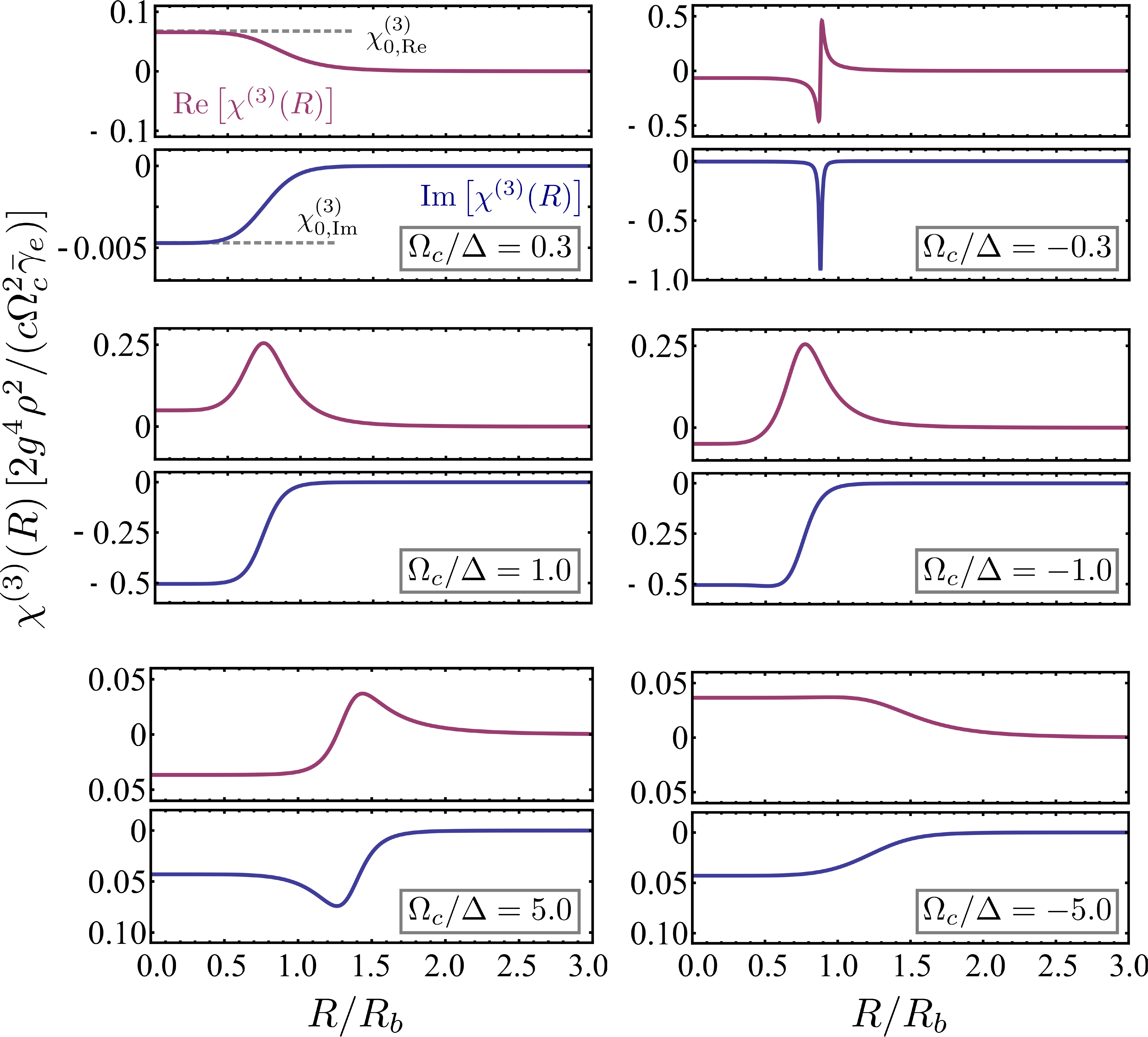}
	\caption{Real (purple) and imaginary part (blue) of the nonlinear susceptibility $\chi^{(3)}(R)$ against the inter-particle distance $R$ for various ratios $\Omega_c/\Delta$ (left column: $\Delta>0$, right column: $\Delta<0$).  The susceptibility is scaled with a factor of $2g^4\rho^2/(c\Omega_c^2\bar{\gamma}_e)$. Plotted with $\Omega_c=2.5\gamma_e$ and $\ket{48\text{S}_{1/2}}$ as the Rydberg state of $^{87}$Rb atoms for a constant atomic density distribution with $\rho=\SI{2e11}{cm^{-3}}$. The blockade radii $R_b$ are $\{3.6,2.9,2.4\}\,\mu \text{m}$ for $\lvert\Omega_c/\Delta\rvert=\{0.3,1.0,5\}$, respectively. }
	\label{fig:chi3_vs_R} 
\end{figure}
After having derived an analytic expression for the nonlinear, nonlocal susceptibility in Eq. (\ref{equ:chi3}), we are in a position to investigate its spatially dependent absorption and refraction features, given by its imaginary and real part, respectively. 

Fig. \ref{fig:chi3_vs_R} displays typical shapes of the nonlinear susceptibility $\chi^{(3)}(R)$ as a function of the inter-atomic distance $R$ for a constant atomic density distribution $\rho(\mathbf{r})=\rho$. For large $R$, the real and imaginary part tend to zero for all ratios $\Omega_c/\Delta$, reflecting the trivial non-interacting regime. 
In the case of $R\rightarrow 0$ the real and imaginary part are constant for a large range of atomic distances $R$ with plateau values $\chi^{(3)}_{0,\text{Re}}$ and $\chi^{(3)}_{0,\text{Im}}$, respectively. For $\Omega_c=\lvert\Delta\rvert$ the latter gets maximal, meaning that the system displays the strongest nonlinear absorption.
 
For intermediate atomic distances $R$ the shape of the real and imaginary part strongly depends on the ratio $\Omega_c/\Delta$ and can display additional features. First, examine the case for a positive single-photon detuning ($\Delta>0$, left column). For $\Omega_c/\Delta <1$, both the imaginary and real part of the nonlinear susceptibility feature a soft-core shape. However, for $\Omega_c=\Delta$ the real part shows a strong maximum and for $\Omega_c>\Delta$ it features a sign-change where the imaginary part gets minimal at a finite distance. For $\Delta<0$ (right column in Fig. \ref{fig:chi3_vs_R}) the situation is reversed, such that the minimum of the imaginary part at a finite distance appears for $\Omega_c <|\Delta|$. Moreover, it is more pronounced than for $\Delta>0$. 

The observed position of the additional features is a direct consequence of the van der Waals interactions and can be understood in terms of the energies of the dressed eigenstates. Examining Fig. \ref{fig:pair-states_energies}(b) we see, that
for $\Delta>0$ (green) the resonance condition is only met for absolute values of the ratio $\Omega_c/\Delta$ being larger than 1, while for $\Delta<0$ (black) the opposite holds. This is exactly the reason, why we observe a minimum of the imaginary part of the nonlinear susceptibility for ratios $\Omega_c/\Delta$ larger (smaller) than 1 for positive (negative) single-photon detunings in Fig. \ref{fig:chi3_vs_R}.

As a result, the ratio $\Omega_c/\Delta$ allows for spatial shaping of the absorption and refraction properties of the nonlinear susceptibility.

\subsection{Scaling of the resonance}
We now discuss the scaling properties of the resonance by looking at the susceptibility for $R\rightarrow 0$.
Assuming $\bar{\gamma}_r=0$ for simplicity we obtain 
\begin{align} 
\chi^{(3)}_{0,\text{Re}} &=\frac{2g^4\rho^2}{c\Omega_c^2}\frac{\Delta\left(\bar{\gamma}_e^2+\Delta^2-\Omega_c^2\right)}{\bar{\gamma}_e^4+\left(\Delta^2-\Omega_c^2\right)^2+2\bar{\gamma}_e^2\left(\Delta^2+\Omega_c^2\right)}\nonumber\\
&\approx
\frac{g^4\rho^2}{2c\Delta^3}\, ,\qquad \text{for } \Omega_c=\lvert\Delta\rvert, \bar{\gamma}_e\ll\lvert\Delta\rvert
\label{equ:susRe_small_R}
\end{align}
for the real part and 
\begin{align} 
\chi^{(3)}_{0,\text{Im}} &=\frac{-2g^4\rho^2}{c\Omega_c^2}\frac{\bar{\gamma}_e\left(\bar{\gamma}_e^2+\Delta^2+\Omega_c^2\right)}{\bar{\gamma}_e^4+\left(\Delta^2-\Omega_c^2\right)^2+2\bar{\gamma}_e^2\left(\Delta^2+\Omega_c^2\right)}\nonumber\\
&\approx
\frac{-g^4\rho^2}{c\bar{\gamma}_e\Delta^2}\, ,\qquad  \text{for } \Omega_c=\lvert\Delta\rvert, \bar{\gamma}_e\ll\lvert\Delta\rvert
\label{equ:susIm_small_R}
\end{align}
for the imaginary part. Here, the second line in Eq. (\ref{equ:susRe_small_R}) and (\ref{equ:susIm_small_R}) gives the value at the resonance condition $\Omega_c=\lvert\Delta\rvert$ in the non-adiabatic limit.

\begin{figure}[t!]
	\includegraphics[width=1\linewidth]{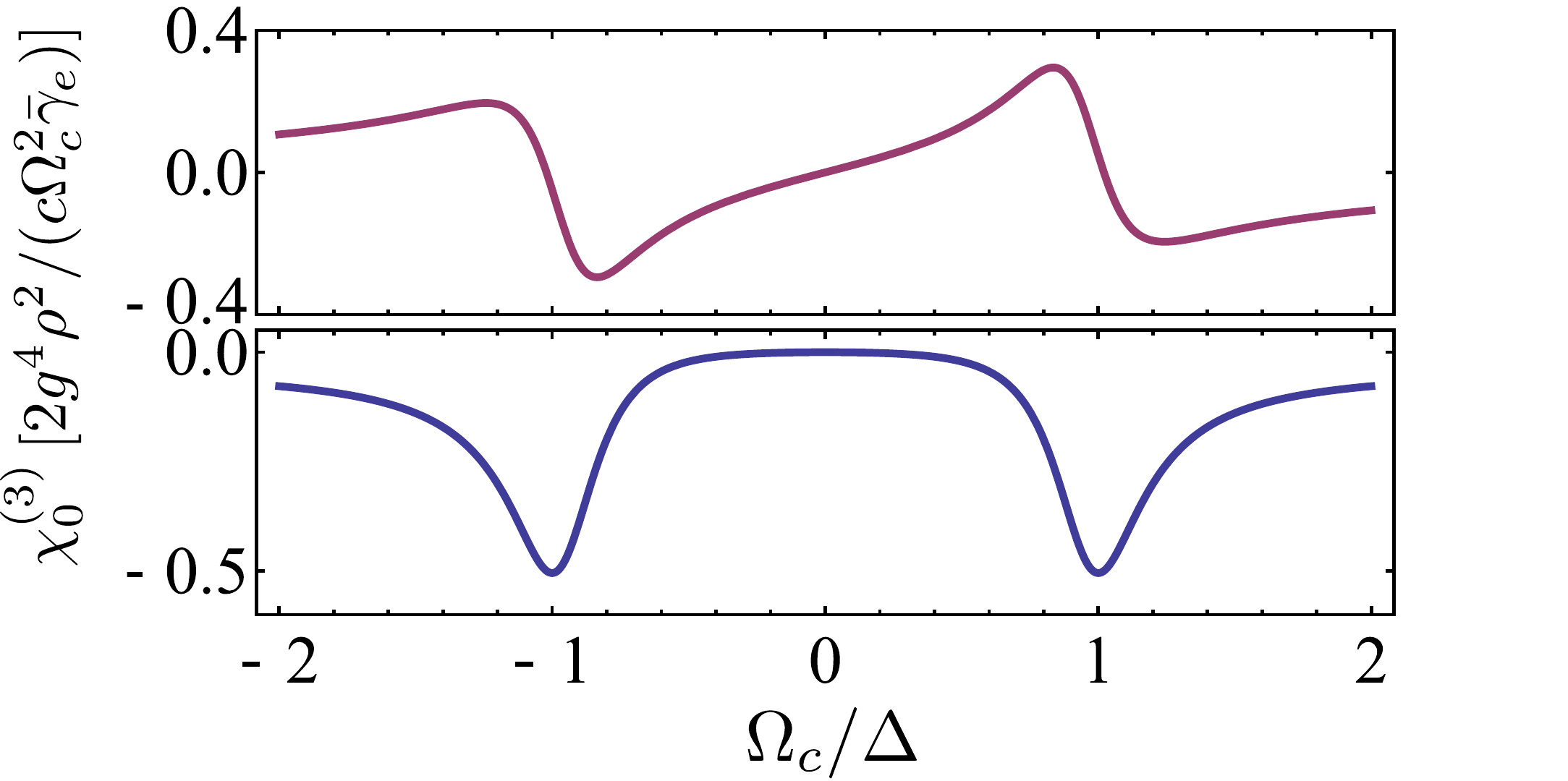}
	\caption{Enhancement of the nonlinear susceptibility. 
		Real (purple) and imaginary part (blue) of the nonlinear susceptibility $\chi_0^{(3)}$ for strong interactions ($V(R)\gg\Omega_c$) against the ratio $\Omega_c/\Delta$.
		At the resonance position ($\Omega_c=\lvert\Delta\rvert$) the real part features a sign change, while the imaginary part is resonantly enhanced. Plotted using $^{87}$Rb atoms with  $\ket{48\text{S}_{1/2}}$ as the Rydberg state and $\Omega_c=2.5\gamma_e$.
	}
	\label{fig:enhancement} 
\end{figure}

Fig. \ref{fig:enhancement}(a) displays the real and imaginary part of  $\chi_0^{(3)}$ as a function of the ratio $\Omega_c/\Delta$. 
Here, the imaginary part is resonantly enhanced for $\Omega_c=\lvert\Delta\rvert$, in agreement with the discussion in the pair-state basis in section \ref{sec:two_body}. The real part exhibits a sign change with a negative slope around $\Omega_c/\Delta=\pm 1$.

At the resonance condition $\Omega_c=\lvert\Delta\rvert$, the imaginary part interestingly depends on the intermediate state decay rate, while the real part does not. This allows to increase the imaginary part independently by choosing an atomic species with a long-lived intermediate state.

\section{Signature of the resonance in the probe transmission}
\label{sec:experiement}
In this section, we investigate whether the two-body, two-photon resonance is experimentally accessible. For this purpose, we solve the propagation Eq. (\ref{equ:propagation_E}). Numerically, this can be done in a straightforward manner by exploiting a split-step Fourier propagation scheme \cite{macnamara2016operator,Hardin:SplitStepFourier:1973}. 

However for a better understanding, we derive an analytic solution of the propagation Eq. (\ref{equ:propagation_E}) under the assumption of a flat input field. Neglecting diffraction, this results in an effective one-dimensional equation \begin{equation}
\partial_z\mathcal{I}(z)=a_1\mathcal{I}(z)+a_2\mathcal{I}^2(z)
\label{equ:propagation_I}
\end{equation}
for the probe field intensity $\mathcal{I}(z)=|\mathcal{E}(z)|^2$, with 
\begin{align}
a_1&=2\text{Im}\left\{\chi^{(1)}\right\},\\
a_2&=2\int\text{d}\mathbf{r'}\text{Im}\left\{\chi^{(3)}(\mathbf{r}-\mathbf{r'})\right\}.\label{equ:a2}
\end{align}
Eq. (\ref{equ:propagation_I}) holds if the probe field intensity is approximately constant over the range of the nonlinear susceptibility (Fig. \ref{fig:chi3_vs_R}). This so called local approximation allows us to reduce the convolution integral in Eq. (\ref{equ:propagation_E}) to an integration solely over the susceptibility in Eq. (\ref{equ:a2}). In addition, we assume in the simplest case a constant atomic density distribution. In this case only $V(\mathbf{r}-\mathbf{r'})$ is left to be position dependent.

A solution of Eq. (\ref{equ:propagation_I}) can be obtained readily and reads
\begin{align}
\mathcal{I}(z)&=\frac{a_1\mathcal{I}_0e^{a_1z}}{a_1+a_2\mathcal{I}_0-a_2\mathcal{I}_0e^{a_1z}}\label{equ:intensity_idealcase}\nonumber\\
&\approx \mathcal{I}_0e^{a_1z}+\frac{a_2}{a_1}e^{a_1z}(e^{a_1z}-1)\mathcal{I}_0^2+\mathcal{O}(\mathcal{I}_0^3),
\end{align}
where the second line is an expansion for a small initial probe field intensity $\mathcal{I}_0=\mathcal{I}(0)$. The first order describes an exponential reduction of the intensity, while the second contains the nonlinear absorption. Eq. (\ref{equ:intensity_idealcase}) provides a leading-order nonlinear description of the probe field's propagation in the limit of a flat input field and a constant intensity distribution of the control field.

\begin{figure}[t!]
	\includegraphics[width=1\linewidth]{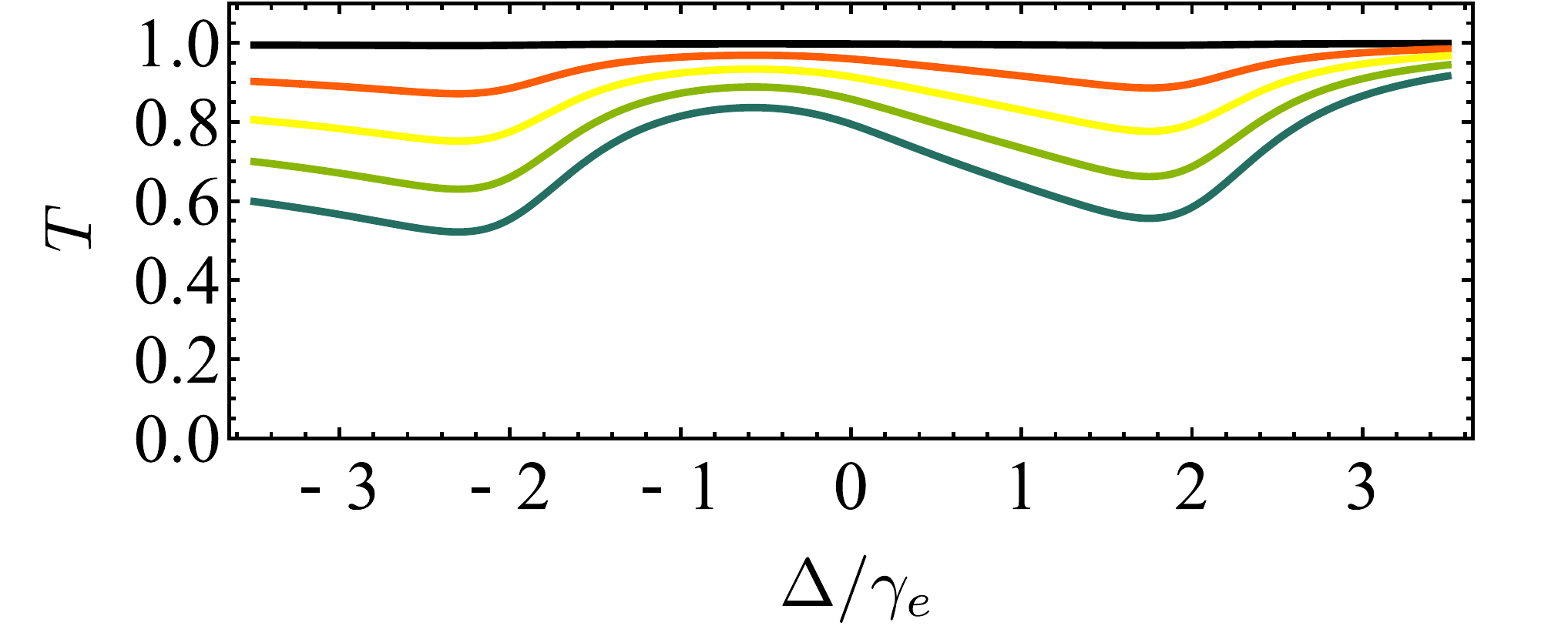}
	\caption{Transmission $T=\mathcal{I}(L)/\mathcal{I}_0$ of the probe field after propagating a distance $L$ as a function of the single-photon detuning $\Delta$ for $\Omega_p/\Omega_c={0.008,0.02,0.04,0.06,0.08}$ (black to blue), respectively. In the nonlinear regime with high $\Omega_p$, two absorption minima for $\Delta\approx\pm\Omega_c$ appear as a consequence of the two-body, two-photon resonance. Plotted for $\{\Omega_c,\delta,\rho,L\} =\{2\gamma_e,\,\, 0,\,\SI{1e11}{cm^{-3}},\, \SI{400}{\mu m}\}$ using Eq. (\ref{equ:intensity_idealcase}).}
	\label{fig:transmission-spectrum} 
\end{figure}

Fig. \ref{fig:transmission-spectrum} shows a transmission spectrum of the probe field as a function of the single-photon detuning. In the non-interacting regime (small $\Omega_p$), the transmission equals 1 for all $\Delta$, due to the EIT effect on two-photon resonance, where $\delta=0$. Increasing the Rabi frequency of the probe field gradually, the interacting, nonlinear regime is reached. Here, two transmission minima occur as a consequence of the enhanced susceptibility at $\Delta\approx\pm\Omega_c$. 

The overall shift of the spectrum towards negative values of $\Delta$ is a result of an integration over the nonlinear susceptibility in Eq. (\ref{equ:a2}), as the shape of its imaginary part exhibits a minimum at a finite distance for a positive (negative) ratio $\Omega_c/\Delta$ above (below) 1, as shown in Fig. \ref{fig:chi3_vs_R}.

The distinct absorption features in the transmission spectrum allow to access the resonance effect experimentally. However, for a realistic experimental situation a Gaussian atomic density distribution should be considered. This is straightforward as explained in Appendix \ref{app-sec:probe_transmission} and the resonance is still observable. The parameters in Fig. \ref{fig:transmission-spectrum}, indicate that the two-body, two-photon resonance is experimentally accessible.

\section{Conclusion and Outlook\label{sec_concl}}
In conclusion, we predicted an enhancement of the nonlinear optical response of an interacting  Rydberg gas under EIT conditions. This enhancement is a consequence of an interaction-induced two-body, two-photon resonance.
We developed a semi-classical theory in the non-adiabatic, many-body regime in order to derive an analytic expression for the nonlinear optical response for arbitrary interaction strengths, non-flat probe fields and non-constant atomic density distributions. We showed the enhancement as well as its scaling properties with relevant field and atom parameters. We demonstrated that the ratio of $\Omega_c/\Delta$ can be used to tune the spatial dependence of the optical response pointing towards prospects of shaping the effective light potential.

In the quantum regime, a sign change of the effective photonic potential has been predicted \cite{Bienias:ScatteringResonances:PRA2014} and indications for an asymmetric behavior of the optical response depending on the sign of the detuning have been reported \cite{Firstenberg:AttractivePhotons:Nature2013,Tiarks:PhaseShift:ScienceAdvances2016}. Our work adds a semi-classical perspective to both.
Moreover, the derived scaling of the enhancement with $1/\gamma_e$ indicates that a highly nonlinear regime could be reached by using atoms with long lived intermediate states as for example Strontium atoms.
Our findings encourage to investigate the yet unexplored non-adiabatic regime of Rydberg-EIT physics for low optical depth per blockade radius.

\subsection*{ACKNOWLEDGMENTS}
This work is part of and supported by the DFG Priority Program "GiRyd 1929" (DFG WE2661/12-1), the DNRF through a Niels Bohr Professorship to T.P., the Heidelberg Center for Quantum Dynamics, the DFG Collaborative Research Center "SFB 1225 (ISOQUANT)", and the European Union H2020 FET flagship project PASQuanS (Grant No.  817482). A.T acknowledges support from the Heidelberg Graduate School for Fundamental Physics (HGSFP). C.H. acknowledges support from the Alexander von Humboldt foundation.

\appendix

\section{Calculation of the probe field transmission with a non-constant atomic density distribution}
\label{app-sec:probe_transmission}
For a realistic experimental situation we consider a Gaussian atomic density distribution. However, we assume that the density is approximately constant in $x,y$-direction resulting in a distribution of the form
\begin{equation}
\rho(z)=\rho_0e^{-z^2/(2\sigma_z^2)},
\label{equ:gaussian_atom_distribution}
\end{equation}
where $\rho_0$ is the peak atomic density. Moreover, we assume that the density is approximately constant over the range of the nonlinear susceptibility (local approximation). Inserting the atomic distribution given by Eq. (\ref{equ:gaussian_atom_distribution}) into Eq. (\ref{equ:propagation_I}) and solving the differential equation results in 
\begin{equation}
T=\frac{\mathcal{I}(z)}{\mathcal{I}_0}=\frac{A(z)}{1-B(z)\mathcal{I}_0}
\label{equ:ana_solution_density}
\end{equation}
for the transmission $T$ of the probe field, where
\begin{align}
A(z) &= \exp\left\{\frac{1}{2}\tilde{a}_1\sqrt{2\pi}\sigma_z\left[1+\text{Erf}\left(\frac{z}{\sqrt{2}\sigma_z}\right)\right]\right\},
\label{equ:ana_solution_density_lin_coeff}\\
B(z) &= \exp\left\{\frac{1}{2}\tilde{a}_1\sqrt{2\pi}\sigma_z\right\}\label{equ:ana_solution_density_integral}\\
&\times\int_{-\infty}^z\diff \xi\, \tilde{a}_2 \exp\left\{\frac{1}{2}\tilde{a}_1\sqrt{2\pi}\sigma_z\text{Erf}\left(\frac{\xi}{\sqrt{2}\sigma_z}\right)-\frac{\xi^2}{\sigma_z^2}\right\} \nonumber
\end{align}
with the error function $\text{Erf}(z)$ and
\begin{align}
\tilde{a}_1&=\frac{a_1}{\exp\left\{ -z^2/(2\sigma_z^2)\right\} }\, ,\nonumber\\
\tilde{a}_2&=\frac{a_2}{\exp\left\{ -z^2/(2\sigma_z^2)\right\}^2 }.
\end{align}

\end{document}